\documentclass[journal]{IEEEtran} %Journal
\usepackage{cite}

\ifCLASSINFOpdf
  \usepackage[pdftex]{graphicx}

  \DeclareGraphicsExtensions{.pdf,.jpeg,.png}
\else
   \usepackage[dvips]{graphicx}

   \DeclareGraphicsExtensions{.eps}
\fi

\usepackage[cmex10]{amsmath}

\usepackage{array}
\usepackage{mdwmath}
\usepackage{mdwtab}

\usepackage{eqparbox}

\usepackage[caption=false,font=footnotesize]{subfig}

\usepackage{multirow}
\usepackage{amsfonts}
\usepackage{amssymb}
\usepackage{rotating}
\usepackage{balance}
\usepackage{enumerate}
\newcommand{\var}{{\mathbb V}\text{ar}}
\newcommand{\sign}{\text{sign}}
\newcommand{\esp}{\mathbb E}

\begin{document}
\title{Optimal Scalar Quantization for Parameter Estimation}
\author{Rodrigo~Cabral Farias*~% 
             and~Jean-Marc~Brossier% <-this % stops a space
\thanks{The authors are with Grenoble Laboratory of Images, Speech, Signal and Automatics, Department of Images and Signals, 38402 Saint Martin d'H\`eres, France (e-mail: rodrigo.cabral-farias@gipsa-lab.grenoble-inp.fr;	  jean-marc.brossier@gipsa-lab.grenoble-inp.fr). This work was partially presented at IEEE ICASSP 2013 \cite{Farias2013c} and it was supported by the Erasmus Mundus EBWII program.}}% <-this % stops a space
%\thanks{Manuscript received January XX, XXXX; revised January XX, XXXX.}}
\markboth{}%
{Shell \MakeLowercase{FARIAS AND BROSSIER: OPTIMAL ASYMPTOTIC QUANTIZATION FOR ESTIMATION}}

%\IEEEpubid{0000--0000/00\$00.00~\copyright~20XX IEEE}


\maketitle

\begin{abstract}
In this paper, we study an asymptotic approximation of the Fisher information for the estimation of a scalar parameter using quantized measurements. We show that, as the number of quantization intervals tends to infinity, the loss of Fisher information induced by quantization decreases exponentially as a function of the number of quantization bits. A characterization of the optimal quantizer through its interval density and an analytical expression for the Fisher information are obtained. A comparison between optimal uniform and non-uniform quantization for the location and scale estimation problems shows that non-uniform quantization is only slightly better. As the optimal quantization intervals are shown to depend on the unknown parameters, by applying adaptive algorithms that jointly estimate the parameter and set the thresholds in the location and scale estimation problems, we show that the asymptotic results can be approximately obtained in practice using only 4 or 5 quantization bits.

\end{abstract}


 %\ifCLASSOPTIONpeerreview
 %\begin{center} \bfseries EDICS Category: SSP-PARE \end{center}
 %\fi
\IEEEpeerreviewmaketitle

%%%%%%%%%%%%%%%%%%%%%%%%%%%%%%%%%%%%%%%%%%%%%%%%%%%%%%%%%%%

\section{Introduction}
\IEEEPARstart{W}{ith} a broad scope of applications ranging from military systems to building and environmental monitoring \cite{Chong2003}, sensor networks have emerged as a domain of signal processing. The shift from a single sensor approach to a multisensor approach, inherent to sensor networks and sensor arrays, introduces new algorithm design problems. Communication rate and complexity constraints have to be considered. 

A common and direct approach to respect these constraints is to quantize the sensor measurements. Although the theory of quantization for measurements reconstruction is extensively studied in the literature \cite{Gray1998}, its extension for the reconstruction of a parameter embedded in the measurements is much less studied. It is clear that the latter of these two subjects is the most interesting in the development of sensing systems and therefore it will be the main subject of this work.

Location parameter estimation based on quantized measurements is studied in \cite{Papadopoulos2001}, where an analysis of the influence of the quantizer input offset on the estimation performance is presented. The same problem is studied in \cite{Ribeiro2006a} in a binary quantization context; more precisely, the problem of setting binary thresholds of multiple sensors to estimate a random parameter is studied in detail. The number of quantization bits is considered to be finite in both works.

High-rate approximations of performance criteria for inference based on quantized measurements are presented in \cite{Poor1988}. Performance for parameter estimation when both the number of quantization bits and the number of measurements are large is detailed in the uniform quantization case.  In \cite{Marano2007} a high-rate approximation for estimation performance is also proposed, in this case the estimated parameter is random and quantization is considered to be scalar but not necessarily uniform. The optimal companding function (quantizer input nonlinear function) and mean squared estimation errors are obtained such that the sum of the quantizer output entropies is minimized, thus giving a characterization of the rate-distortion function for Bayesian estimation in the asymptotic regime. 

This work is focused on the analysis of asymptotic non-uniform scalar quantization for the estimation of a scalar parameter (asymptotic here means that both the number of samples and quantization bits per sample are large), thus extending the results for uniform quantization presented in \cite{Poor1988}. Differently from \cite{Marano2007}, the parameter is considered to be deterministic and an approximation of the Fisher information (FI) will be written as a function of a density of quantization intervals (the point density). The optimal interval densities and an asymptotic analytical approximation of the FI will then be obtained. The results will be applied to location and scale\footnote{We use a scale parameter characterization of the distribution instead of a standard deviation characterization, to include distributions for which the standard deviation does not exist, \textit{e.g.} the Cauchy distribution.} estimation problems for generalized Gaussian and Student-t distributions. For specific members of these distributions, the Gaussian distribution and the Cauchy distribution, we present a comparison between the theoretical approximation of the asymptotic maximum FI, the FI with optimal uniform quantization and the FI for a practical approximation of the optimal interval density.

Finally, as the practical approximation of the optimal quantizer is shown to depend on the unknown parameter that is estimated, adaptive algorithms that jointly estimate the parameter and set the quantizer thresholds are presented as a solution to achieve asymptotically optimal quantization for estimation. First, we present an adaptive algorithm to estimate a location parameter, then we present its corresponding version to estimate a scale parameter and, at the end, we blend the two solutions to obtain an adaptive algorithm that is asymptotically optimal for the estimation of a location parameter, even if the scale is unknown.

\section{ASYMPTOTIC APPROXIMATION}
\label{sec:p_state}
We consider the problem of estimation of a scalar deterministic parameter $ x\in\mathbb{R} $ of a continuous distribution based on $ N $ independent measurements $ \mathbf{Y}=\left[ Y_{1} \; Y_{2}\; \cdots\; Y_{N}\right]^{T}  $ from this distribution.  Due to the context explained above, estimation of $ x $ will not be done directly based on the continuous amplitude measurements $ \mathbf{Y} $, instead it will be based on a scalar quantized version of $ \mathbf{Y} $, that we denote 
\begin{equation}
\mathbf{i}= \left[ i_{1} \; i_{2}\; \cdots\; i_{N}\right]^{T}= \left[Q\left( Y_{1}\right)  \; Q\left( Y_{2}\right) \; \cdots\; Q\left( Y_{N}\right) \right]^{T},\nonumber
\label{eq1}
\end{equation}
\noindent $ Q $ represents the scalar quantizer and it is given by
\begin{equation}
Q\left(Y\right)=i , \quad \text{if} \quad Y\in q_{i}=\left[\tau_{i-1},\tau_{i}\right),
\label{eq2}
\end{equation}
\noindent where $ i\in\left\lbrace 1, \cdots, N_{I} \right\rbrace  $, $ N_{I} $ is the number of quantization intervals $ q_{i} $ and $ \tau_{i} $ are the quantizer thresholds. We denote the first and last thresholds $ \tau_{0}=\tau_{\text{min}} $ and $ \tau_{N_{I}}=\tau_{\text{max}} $.

A lower bound on the variance of any unbiased estimator $ \hat{X} $ of $ x $ based on $ \mathbf{i} $ is the Cram\'er--Rao bound (CRB). Under the independence assumption and supposing that the support of the marginal cumulative distribution function (CDF) $ F $ of the continuous measurements does not depend on $ x $, the CRB is
\begin{equation}
\var\left[\hat{X}\right]\geq CRB_{q} =\frac{1}{N I_{q}} ,
\label{eq3}
\end{equation}
\noindent where $ I_{q} $ is the FI for a quantized measurement, which can be written as
\begin{eqnarray}
I_{q}&=&\esp\left[S_{q}^{2}\right]=\esp\left\lbrace \left[ \frac{\partial\log\mathbb{P}\left(i;x\right) }{\partial x}\right] ^{2}\right\rbrace \nonumber\\
       &=&\sum\limits_{i=1}^{N_{I}} \left[ \frac{\partial\log\mathbb{P}\left(i;x\right)  }{\partial x}\right] ^{2}\mathbb{P}\left(i;x\right),
\label{eq4}
\end{eqnarray}
\noindent $ S_{q} $ is the score function for quantized measurements and $ \mathbb{P}\left(i;x\right) $ is the probability of having the quantizer output $ i $:
\begin{equation}
\mathbb{P}\left(i;x\right)=F\left(\tau_{i};x\right)-F\left(\tau_{i-1};x\right).
\label{eq5}
\end{equation}
\noindent Under some regularity assumptions the CRB is known to be asymptotically ($ N\rightarrow+\infty $) tight and it can be attained asymptotically using maximum likelihood estimation (MLE). Thus, we can approximately assess estimation performance through the CRB and consequently through the FI.

Note that the FI in (\ref{eq4}) depends not only on the CDF of the measurement distribution $ F $ and on the parameter $ x $, but also on the quantizer thresholds (or equivalently on the quantizer intervals). Therefore, two questions naturally arise: 
\begin{enumerate}[i)]
\item what are the optimal thresholds maximizing the FI?
\item For these optimal thresholds, what is the value of the FI as a function of $ N_{I} $?
\end{enumerate}
  
These questions can be easily answered only in a few cases. In binary quantization the first question can be answered by maximizing the FI through the choice of the (single) threshold position. This is a one dimensional problem which can be solved through exhaustive search. 

In uniform quantization, an answer to the first question can be obtained through the maximization of the FI with respect to (w.r.t.) the central quantizer position and the quantization interval length. The second question can be trivially answered with the solution of the first.

For a non-uniform quantizer, maximization of $ I_{q} $ w.r.t. $ \tau_{i} $ or $ q_{i} $ is a difficult problem. 

To answer these questions, we chose to follow the asymptotic approach (well-known from standard quantization for measurement reconstruction). That is to say, instead of keeping the number of quantizer intervals small to solve the maximization problem with exhaustive search, we will consider that $ N_{I} $ is large and the quantizations interval lengths $ \Delta_{i} $ are small, so that we can have an analytical approximation of $ I_{q} $. In a first view, this seems to be in contradiction with the context, which imposes small $ N_{I} $. Fortunately, we will see that such contradiction does not exist in practice since the asymptotic results can be attained even for small number of quantization bits ($ 4 $ and $ 5 $).

We assume that $ F\left(y;x\right)  $ admits a probability density function (PDF) $ f\left(y;x\right) $ which is positive, smooth in both $ y $ and $ x $ and defined on a bounded support. Following the development presented in \cite{Marano2007}, the quantity $ \esp\left[\left(S_{c}-S_{q}\right)^{2} \right]  $ (with $ S_{c}=\frac{\partial\log f\left(y;x\right) }{\partial x} $ the score function for estimation based on continuous measurements) can be analyzed to obtain a characterization of $ I_{q} $, we can rewrite  $ \esp\left[\left(S_{c}-S_{q}\right)^{2} \right]  $ as
\begin{equation}
\esp\left[\left(S_{c}-S_{q}\right)^{2} \right]=I_{c}+I_{q}-2\esp\left[S_{c}S_{q}\right],  
\label{eq7}
\end{equation}
\noindent $ I_{c} $ is the FI for continuous measurements. The definitions of the quantities $ S_{c} $ and $ S_{q} $ can be used to obtain $ \esp\left[S_{c}S_{q}\right] = \esp\left[S_{q}^{2}\right]  $. Thus,
\begin{equation}
I_{q}=I_{c}-\esp\left[\left(S_{c}-S_{q}\right)^{2} \right].
\label{eq8}
\end{equation}
\noindent This expression shows that quantization cannot increase the FI as the term on the right can be interpreted as a loss of performance due to quantization.

\paragraph*{Remark}as the loss is positive, we can see that $ I_{q} $ is bounded above by $ I_{c} $. Also, if we add a threshold to a quantizer, by the data processing inequality for the Fisher information \cite{Zamir1998}, we know that the FI for the new quantizer will be equal to or larger than the FI for the quantizer without the new threshold. Thus, as the FI is bounded above, a sequence of quantizers with increasing number of thresholds will have a FI that converges to a constant. The question that arises here is the following: can we make $ I_{q} $ tend to $ I_{c} $? We will see in what follows that this is possible.

The loss due to quantization, which we denote $ L $, must be minimized with respect to the quantization intervals. $ L $, which is an expectation, can be rewritten as a sum of integrals, each term of the sum represents the loss produced in a quantization interval:
\begin{equation}
L=\sum\limits_{i=1}^{N_{I}}\int\limits_{q_{i}} \left[\frac{\partial\log f\left(y;x\right) }{\partial x}-\frac{\partial\log\mathbb{P}\left(i;x\right)  }{\partial x} \right]^{2}f\left(y;x\right)\text{d}y .
\label{eq9}
\end{equation}

\subsection*{First term $ \frac{\partial\log f\left(y;x\right) }{\partial x} $}

For the quantization interval $ q_{i} $, we can approximate the PDF with a Taylor series around the quantization interval central point $ y_{i}=\frac{\tau_{i}+\tau_{i-1}}{2} $:
\begin{equation}
f\left(y;x\right)= f_{i}+f_{i}^{\left(y\right)} \left(y-y_{i}\right)+\frac{f_{i}^{\left(yy\right)}}{2}\left(y-y_{i}\right)^{2}+o\left(y-y_{i}\right)^{2},
\label{eq10}
\end{equation}
\noindent the variables for which the function is differentiated are indicated by the superscripts. The subscript represents function evaluation (after differentiation) at $ y_{i} $. We will assume that the sequences of intervals for increasing $ N_{I} $ are chosen such that for any $ \varepsilon>0 $ it is possible to find a $ N_{I}^{*} $ for which
\begin{equation}
\frac{o\left(y-y_{i}\right)^{2}}{\left(y-y_{i}\right)^{2}}<\varepsilon,\quad\text{for}\,N_{I}> N_{I}^{*},\,y\in q_{i}.
\label{eq11}
\end{equation}
\noindent As $ f>0 $, $\log f $ at interval $ q_{i} $ can be approximated also using a Taylor expansion:
\begin{eqnarray}
\log f\left(y;x\right) = \log f_{i}+\left(\log f \right)_{i}^{\left(y\right) }\left(y-y_{i}\right)+\nonumber\\
+\left(\log f \right)_{i}^{\left(yy\right) }\frac{\left(y-y_{i}\right)^{2}}{2} +o\left(y-y_{i}\right)^{2}
\label{eq12}
\end{eqnarray}
\noindent and its derivative w.r.t. $ x $ is
\begin{eqnarray}
\frac{\partial \log f\left(y;x\right)}{\partial x} =\left(\log f\right)_{i}^{\left(x\right) }+\left(\log f\right)_{i}^{\left( yx\right) }\left(y-y_{i}\right) +\nonumber\\
+\left(\log f \right)_{i}^{\left(yyx\right) }\frac{\left(y-y_{i}\right)^{2}}{2} +o\left(y-y_{i}\right)^{2}.
\label{eq13}
\end{eqnarray}

\subsection*{Second term $ \frac{\partial\log\mathbb{P}\left(i;x\right)  }{\partial x} $}

Now, we calculate the other term in the squared factor. Integrating the PDF in (\ref{eq10}) on the interval $ q_{i} $ with step-length $ \Delta_{i} $, we get
\begin{equation}
\mathbb{P}\left(i,x\right)=f_{i}\Delta_{i}+f_{i}^{\left(yy\right)}\frac{\Delta_{i}^{3}}{24}+o\left(\Delta_{i}^{3}\right). 
\label{eq14}
\end{equation}
\noindent Note that the term in $ \Delta_{i}^{2} $ is zero since $ y_{i} $ is the interval central point and the integral of $ \left( y-y_{i}\right)  $ around it is zero. The logarithm of $ \mathbb{P}\left(i,x\right) $ can be obtained by dividing the second and third terms of the right hand side of (\ref{eq14}) by the first term and then using the Taylor series for $ \log\left(1+x\right)=x+\circ\left(x\right)   $. Differentiating the resulting expression w.r.t. $ x $ gives
\begin{equation}
\frac{\partial \log \mathbb{P}\left(i,x\right)}{\partial x}= \left(\log f\right)_{i}^{\left(x\right)}+\left(\frac{f^{\left(yy\right) }}{f}\right)_{i}^{\left(x\right)}\frac{\Delta_{i}^{2}}{24}+o\left(\Delta_{i}^{2}\right).
\label{eq15}
\end{equation}

\subsection*{Asymptotic expression for the loss $ L $}

Subtracting (\ref{eq15}) from (\ref{eq13}) and squaring makes the leading term with least power in $ \left(y-y_{i} \right)  $ or in $ \Delta_{i} $ to be $ \left(\log f\right)_{i}^{\left( yx\right) }\left(y-y_{i}\right) $. When we square this difference and multiply by the Taylor series of $ f $, we have a leading term $ \left[ \left(\log f\right)_{i}^{\left( yx\right) }\right]^{2}f_{i}\left(y-y_{i}\right)^{2} $ and all other terms have larger powers of $ \left(y-y_{i}\right) $ and/or $ \Delta_{i} $. Therefore, after integrating the squared difference multiplied by the Taylor series of $ f $, we get
\begin{eqnarray}
L&=&\sum\limits_{i=1}^{N_{I}}\left\lbrace \left[ \left(\log f\right)_{i}^{\left( yx\right) }\right] ^{2}f_{i}\frac{\Delta_{i}^{3}}{12}+o\left(\Delta_{i}^{3}\right)\right\rbrace\nonumber\\
&=&\sum\limits_{i=1}^{N_{I}}\left\lbrace \left( S_{c,i}^{\left( y\right) }\right)^{2}f_{i}\frac{\Delta_{i}^{3}}{12}+o\left(\Delta_{i}^{3}\right)  \right\rbrace .
\label{eq16}
\end{eqnarray}
\noindent To obtain a characterization of the loss w.r.t. the quantizer, an interval density function $ \lambda\left(y\right) $ will be defined:
\begin{equation}
\lambda\left(y\right)=\lambda_{i}=\frac{1}{N_{I}\Delta_{i}},\quad\text{for } y\in q_{i}.
\label{eq17}
\end{equation}
The interval density when integrated in an interval gives, roughly, the fraction of the number of quantization intervals contained in that interval. It is a positive function that always sums to one (the Riemann sum is equal to one $ \sum\limits_{i=1}^{N_{I}} \frac{1}{N_{I}\Delta_{i}}\Delta_{i}=1\approx\int \lambda\left(y\right)\text{d}y. $).

\noindent Rewriting (\ref{eq16}) with the interval density gives
\begin{equation}
L=\sum\limits_{i=1}^{N_{I}}\left\lbrace \left( S_{c,i}^{\left( y\right) }\right)^{2}f_{i}\frac{\Delta_{i}}{12 N_{I}^{2}\lambda_{i}^{2}}+o\left(\frac{1}{N_{I}^{2}}\right)\Delta_{i}  \right\rbrace .
\label{eq18}
\end{equation}
\noindent We suppose that all $ \Delta_{i} $ converge uniformly to zero as $ N_{I}\rightarrow\infty $. As a consequence,
\begin{equation}
\underset{N_{I}\rightarrow\infty}{\lim}N_{I}^{2}L=\frac{1}{12}\int\frac{\left( \frac{\partial S_{c}\left(y;x\right)  }{\partial y}\right)^{2}f\left(y;x\right) }{\lambda^{2}\left(y\right)}\text{d}y.
\label{eq19}
\end{equation}

\subsection*{Approximation of the FI $ I_{q} $}
\noindent Expression (\ref{eq19}) indicates that, for $ N_{I} $ large enough, the following approximation of the FI can be used
\begin{equation}
I_{q}\approx I_{c}-\frac{1}{12 N_{I}^{2}}\int\frac{\left( \frac{\partial S_{c}\left(y;x\right)  }{\partial y}\right)^{2}f\left(y;x\right) }{\lambda^{2}\left(y\right)}\text{d}y.
\label{eq20}
\end{equation}
\noindent This expression shows that:
\begin{itemize}
\item in the asymptotic regime, if the quantizer intervals are chosen in a way such that all $ \Delta_{i} $ tend to zero uniformly, then the asymptotic estimation performance for quantized measurements will tend to the estimation performance for continuous measurements. This answer the question on the convergence of $ I_{q} $.
\item The convergence is quadratic in $ N_{I} $, or equivalently exponential in the number of bits $ N_{B}=\log_{2}\left(N_{I}\right)  $ as 
\begin{equation}
I_{q}\approx I_{c}-\frac{2^{-2N_{B}}}{12}\int\frac{\left( \frac{\partial S_{c}\left(y;x\right)  }{\partial y}\right)^{2}f\left(y;x\right) }{\lambda^{2}\left(y\right)}\text{d}y.\nonumber
\end{equation}
\item Maximization of the estimation performance through the quantizer can be done by minimizing the integral in (\ref{eq20}) w.r.t. $ \lambda\left(y\right)  $. This leads to an asymptotic characterization of the optimal quantizer for estimation.
\end{itemize}

\subsection*{Optimal interval density $ \lambda^{\star} $ and maximum FI $ I_{q}^{\star} $}
The optimal interval density $ \lambda^{\star} $ maximizing (\ref{eq20}) and the corresponding maximum FI $ I_{q}^{\star} $ can be obtained by applying the H\"older's inequality to the integral. We obtain
\begin{equation}
\lambda^{\star}\left(y\right)=\frac{ \left( \frac{\partial S_{c}\left(y;x\right)  }{\partial y}\right)^{\frac{2}{3}}f^{\frac{1}{3}}\left(y;x\right) }{\int \left( \frac{\partial S_{c}\left(y;x\right)  }{\partial y}\right)^{\frac{2}{3}}f^{\frac{1}{3}}\left(y;x\right)\text{d}y},
\label{eq21}
\end{equation}
\begin{equation}
I_{q}^{\star}\approx I_{c}-\frac{2^{-2N_{B}}}{12}\left[\int\left( \frac{\partial S_{c}\left(y;x\right)  }{\partial y}\right)^{\frac{2}{3}}f^{\frac{1}{3}}\left(y;x\right)\text{d}y \right]^{3}.
\label{eq22}
\end{equation}
\noindent This expression is the answer for the last question which was still unanswered. Notice that $ f^{\frac{1}{3}} $ from standard quantization (Panter and Dite approximation \cite{Panter1951}) is present in this expression, however, a multiplying factor depending on the sensibility of the score function for the estimation problem is also present. As a consequence, all estimation problems for which the sensibility of the score function is not constant will have optimal quantizers for estimation that differ from those for reconstruction of the continuous measurements.

A practical way of approximating the optimal thresholds can be obtained if we use the definition of the interval density: the percentage of intervals until the interval $ q_{i} $, $ \frac{i}{N_{I}} $ must equal the integral of the interval density from $ \tau_{\text{min}} $ to $ \tau_{i} $. This leads to the following approximation of the optimal thresholds
\begin{equation}
\tau_{i}^{\star}=F_{\lambda}^{-1}\left(\frac{i}{N_{I}}\right),
\label{eq29}
\end{equation}
\noindent where $ F_{\lambda}^{-1}\left(\cdot\right)  $ is the inverse of the CDF related to $ \lambda^{\star} $.

\section{Application of the results}
We apply the results above to obtain the optimal interval densities, the maximum FI approximation and the practical approximation for the optimal thresholds in two estimation problems:
\begin{itemize}
\item \emph{Location estimation: }the objective is to estimate a parameter $ x=\mu $ of a CDF $ F\left(\frac{y-\mu}{\delta} \right)  $. The median of $ Y_{k} $ is zero when $ \mu $ is zero, therefore, $ \mu $ represents the median of the distribution. For symmetric and unimodal distributions where the mean exists, \textit{e.g.} the Gaussian distribution, the location parameter is either the mode, the median and the mean of the distribution.
\item \emph{Scale estimation: }the parameter to be estimated is $x=\delta $ in the previous CDF, which is constrained to be strictly positive. When the standard deviation exists this parameter is proportional to it.
\end{itemize}

We focus on two families of distributions: 
\begin{itemize}
\item \emph{generalized Gaussian distributions (GGD): }these distributions are used in signal processing to model noise which is not necessarily Gaussian \cite{Varanasi1989}. Their PDF are given by\footnote{$\Gamma\left(x\right)=\int\limits_{0}^{+\infty} w^{x-1} \exp\left(-w \right)\,\text{d}w$ is the Gamma function.} 
\begin{equation}
f_{GGD}\left(y;\mu,\delta\right) = \frac{\beta}{2\delta\Gamma\left(\frac{1}{\beta}\right) }\exp\left( -\left\vert \frac{y-\mu}{\delta} \right\vert^{\beta}\right) , \label{eq_pdf_ggd}
\end{equation}
\noindent where $ \beta $ is a strictly positive shape parameter. 
\item \emph{Student t distributions (STD): }the motivation for considering Student-t distributions comes from its use in robust statistics \cite{Lange1989}. Their PDF are the following\footnote{$  B\left(x,y\right)=\int\limits_{0}^{1} w^{x-1} \left(1-w\right)^{y-1} \,\text{d}w$ is the Beta function.}
\begin{equation}
f_{STD}\left(y;\mu,\delta\right) = \frac{ \left[ 1+\frac{1}{\beta}\left( \frac{y-\mu}{\delta}\right) ^{2}\right] ^{-\frac{\beta +1}{2}} }{\delta\sqrt{\beta} B\left(\frac{\beta}{2},\frac{1}{2}\right) } . \label{eq_pdf_std}
\end{equation}
\noindent We will see later that even if we cannot, in general, obtain analytic expressions for the the optimal thresholds and for the approximation of the maximum FI in the location parameter case, we can find expressions for its most commonly used member, the Cauchy distribution ($ \beta=1 $) and we can also find analytic expressions for the scale estimation problem in general.
\end{itemize}

Observe that the support of these distributions is unbounded, thus they do not respect one of the hypothesis stated above. As in standard quantization theory, we expect that the error produced by neglecting the effects of the overload region (the extremal regions) will be small.

\subsection{Location parameter estimation}
\label{subsec:loc_par_est}

\subsubsection*{Generalized Gaussian distribution}
the first step to obtain the practical approximation of the optimal thresholds is to evaluate the optimal interval density (\ref{eq21}). Using the expression of the GGD PDF (\ref{eq_pdf_ggd}), we can see that the derivative of the score does not exist at $ y=\mu $ for $ \beta\leq 1 $, thus we proceed only for $ \beta>1 $. The following interval density can be obtained:
\begin{equation}
\lambda_{GGD}^{\star,\mu}\left(y\right) = \frac{\left\vert \frac{y-\mu}{\delta} \right\vert^{\frac{2\beta-4}{3}}\exp\left(-\frac{1}{3}\left\vert \frac{y-\mu}{\delta} \right\vert^{\beta} \right) }{C_{GGD}^{\mu}},
\label{eq_lambda_ggd_mu}
\end{equation}
\noindent where $ C_{GGD}^{\mu} $ is a constant normalizing the density. Using the symmetry of the density around $ \mu $ and the change of variables $ \varepsilon=\frac{1}{3}\left(  \frac{y-x}{\delta} \right) ^{\beta} $ we can get an expression\footnote{$ \gamma\left(x,y \right)= \int\limits_{0}^{y} w^{x-1} \exp\left(-w \right)\,\text{d}w $ is the incomplete Gamma function.} for the CDF related to $ \lambda^{\star} $:
\begin{equation}
F_{\lambda,GGD}^{\mu}\left(y\right)  = \frac{1}{2}+\frac{\sign\left(y-\mu\right) }{2}\frac{\gamma\left[\frac{1}{3}\left(2-\frac{1}{\beta}\right) ,\frac{1}{3}\left\vert \frac{y-\mu}{\delta} \right\vert^{\beta} \right]}{\Gamma\left[\frac{1}{3}\left(2-\frac{1}{\beta}\right) \right]}.\nonumber
\end{equation}
\noindent Using the inverse of this function we obtain the approximately optimal thresholds (\ref{eq29}). For $ i\in\left\lbrace 1,\;\cdots,\;N_{I} \right\rbrace  $
\begin{equation}
\tau_{i,GGD}^{\star,\mu}=\mu+\delta\sign\left(\frac{2i}{N_{I}}-1\right)\alpha_{GGD}^{\mu}\left( \left\vert \frac{2i}{N_{I}}-1 \right\vert \right) ,
\label{eq_tau_mu_ggd}
\end{equation}
\noindent with
\begin{equation}
\alpha_{GGD}^{\mu}\left(x\right)= \left\lbrace 3\gamma^{-1}\left\lbrace  \frac{1}{3}\left(2-\frac{1}{\beta}\right),x\Gamma\left[\frac{1}{3}\left(2-\frac{1}{\beta}\right) \right]\right\rbrace \right\rbrace^{\frac{1}{\beta}}  ,
\label{eq_alpha_mu_ggd}
\end{equation}
\noindent where $ g^{-1}\left(\cdot\right) $ denotes the inverse of the function $ g\left(\cdot\right) $.

\paragraph*{Remarks} 
\begin{enumerate}[i)]
\item for the Gaussian distribution, the derivative of the score is constant, thus the optimal interval density is proportional to $ f^{\frac{1}{3}} $ and optimal quantization for estimation is asymptotically equivalent to optimal quantization for reconstruction of the continuous measurements $ Y_{k} $.
\item Notice that the distribution with largest $ \beta $ for which the evaluation of $ \lambda^{\star} $ cannot be done is the Laplacian distribution ($ \beta=1 $), this might be related to the fact that binary quantization with a central threshold placed at $ \mu $ is optimal in the context of estimation, as in this case $ I_{q}=\frac{1}{\delta^{2}}=I_{c} $.  
\item Note also that the distributions with $ \beta>2 $ have a maximum of $ \lambda^{\star} $ at a point different from $ \mu $, thus the point where most of the statistical information is located is not around the true value of the parameter but it is shifted from it. This can be linked to the fact that optimal binary quantization for this subclass of the GGD must be performed by placing the threshold at a point different from $ \mu $. As for $ \beta>2 $ the PDF is flat around zero, similarly to the uniform distribution, the points where most of the information is located are in the border of the flat region.
\end{enumerate}

\begin{table*}[t!]
\centering
\begin{tabular}{ c || c c c || c c c}
 \multicolumn{1}{c}{} & \multicolumn{3}{c}{Gaussian ($ I_{c}=2 $)}&\multicolumn{3}{c}{Cauchy ($ I_{c}=0.5 $)}\\\hline
$ N_{B}  $&Optimal&Uniform&\parbox[c]{1.2cm}{Practical\\approx.}&Optimal&Uniform&\parbox[c]{1.2cm}{\vspace{2pt} Practical \\ approx. \vspace{2pt}}\\\hline\hline
$ 1 $&$1.27323954^{\dagger}$&$1.27323954$&$1.27323954$&$0.40528473^{\dagger}$&$0.40528473$&$0.40528473$\\\hline
$ 2 $&$1.76503630^{\dagger}$&$1.76503630$&$1.75128300$&$0.43433896^{\dagger}$&$0.43433896$&$0.40528473$\\\hline
$ 3 $&$1.93090199^{\dagger}$&$1.92837814$&$1.92740111$&$0.48474865^{\dagger}$&$0.45600797$&$0.47893785$\\\hline
$ 4 $&$1.97874454^{\star}$&$1.97841622$&$1.98038526$&$0.49533850^{\star}$&$0.48136612$&$0.49504170$\\\hline
$ 5 $&$1.99468613^{\star}$&$1.99353005$&$1.99489906$&$0.49883463^{\star}$&$0.49204506$&$0.49879785$\\\hline
$ 6 $&$1.99867153^{\star}$&$1.99807736$&$1.99869886$&$0.49970866^{\star}$&$0.49656712$&$0.49970408$\\\hline
$ 7 $&$1.99966788^{\star}$&$1.99943563$&$1.99967136$&$0.49992716^{\star}$&$0.49851056$&$0.49992659$\\\hline
$ 8 $&$1.99991697^{\star}$&$1.99983649$&$1.99991741$&$0.49998179^{\star}$&$0.49935225$&$0.49998172$\\\hline
\end{tabular}
\caption{Fisher information (FI) for the estimation of location parameters of Gaussian and Cauchy distributions based on quantized measurements. $ N_{B} $ is the number of quantization bits. $ \textit{Optimal}^{\dagger} $ shows the FI obtained by exhaustive search of the optimal thresholds. $ \textit{Optimal}^{\star} $ is the asymptotic approximation of the FI. \textit{Uniform} shows the FI for optimal uniform quantization and \textit{Practical approx.} gives the FI for the practical approximation of the optimal thresholds for estimation. }
\label{tab1}
\end{table*}

Now we can approximate the maximum FI. In this case, $ I_{c} $ is
\begin{equation}
I_{c,GGD}^{\mu}=\frac{1}{\delta^{2}}\frac{\beta\left(\beta-1\right)\Gamma\left( 1-\frac{1}{\beta}\right)  }{\Gamma\left(\frac{1}{\beta}\right) }.
\label{ic_ggd}
\end{equation}

Using (\ref{ic_ggd}) in (\ref{eq22}) and evaluating the integral term, we have
\begin{eqnarray}
I_{q,GGD}^{\star,\mu}\approx\frac{1}{\delta^{2}}\frac{\beta-1 }{\Gamma\left(\frac{1}{\beta}\right) }\left\lbrace \beta \Gamma\left(1-\frac{1}{\beta} \right)-\right.\hspace{40pt}\\
\left.2^{-2N_{B}}\left(\beta-1\right)3^{1-\frac{1}{\beta}}\Gamma^{3}\left[\frac{1}{3}\left(2-\frac{1}{\beta} \right)  \right]    \right\rbrace. \hspace{-25pt}\nonumber
\label{iq_ggd}
\end{eqnarray}

In the Gaussian case ($ \beta=2 $), this approximation is 
\begin{equation}
I_{q,G}^{\star,\mu}\approx\frac{2}{\delta^{2}}\left[1-\pi\sqrt{3}\,2^{-\left(2N_{B}+1\right)}\right].
\label{eq26}
\end{equation}

\subsubsection*{Student t distribution (Cauchy distribution)} the interval density can be evaluated using (\ref{eq_pdf_std}) in (\ref{eq21}), which gives
\begin{equation}
\lambda_{STD}^{\star,\mu}\left(y\right)=\frac{1}{C_{STD}^{\mu}}\frac{\left[ 1-\frac{1}{\beta}\left( \frac{y-\mu}{\delta}\right) ^{2}\right]^{\frac{2}{3}} }{\left[1+\frac{1}{\beta}\left( \frac{y-\mu}{\delta}\right) ^{2} \right]^{\frac{9+\beta}{6}}}.
\label{eq_lambda_std_mu}
\end{equation}
\noindent Unfortunately the CDF of this density cannot be expressed analytically (with known special functions). Thus, for obtaining an expression for the optimal thresholds, it is necessary to numerically integrate the density and then invert an interpolation of the numerical integration.

In the case of the Cauchy distribution ($ \beta=1 $), we can evaluate analytically the CDF. Integration of $ \lambda^{\star} $ and inversion (we omit here the CDF expression) with $ i'= i-\frac{N_{I}}{2}  $ leads to\footnote{$ I_{z}\left(x,y\right)=\int\limits_{0}^{z} w^{x-1} \left(1-w\right)^{y-1} \,\text{d}w $ is the incomplete Beta function.}
\begin{equation}
\tau_{i,C}^{\star,\mu}=
\begin{cases}
\mu+\delta\sign\left(i' \right)\sqrt{ \frac{ 1-\sqrt{I_{B\left(\frac{1}{2},\frac{5}{6} \right)\left( 1- \frac{ 4\left\vert i' \right\vert }{ N_{I} } \right) }^{-1}\left(\frac{1}{2},\frac{5}{6} \right) }}{1+\sqrt{I_{B\left(\frac{1}{2},\frac{5}{6} \right)\left( 1- \frac{ 4\left\vert i' \right\vert }{ N_{I} } \right) }^{-1}\left(\frac{1}{2},\frac{5}{6} \right) }}},\\ \qquad\qquad\qquad\qquad\qquad\qquad\qquad\text{when }\left\vert i'\right\vert\leq\frac{1}{4}, \\
\mu+\delta\sign\left(i' \right)\sqrt{ \frac{ 1+\sqrt{I_{B\left(\frac{1}{2},\frac{5}{6} \right)\left( \frac{ 4\left\vert i' \right\vert }{ N_{I} }-1 \right) }^{-1}\left(\frac{1}{2},\frac{5}{6} \right) }}{1-\sqrt{I_{B\left(\frac{1}{2},\frac{5}{6} \right)\left( \frac{ 4\left\vert i' \right\vert }{ N_{I} }-1 \right) }^{-1}\left(\frac{1}{2},\frac{5}{6} \right) }}},\\ \qquad\qquad\qquad\qquad\qquad\qquad\qquad\text{when }\left\vert i'\right\vert\leq\frac{1}{4}. \\
\end{cases}\label{eq_tau_mu_cauchy}
\end{equation}

The continuous measurement FI for this distribution is $ I_{c,C}^{\mu}=\frac{1}{2}\frac{1}{\delta^{2}} $. Evaluating $ I_{q,C}^{\star,\mu} $ (\ref{eq22}), we have
\begin{equation}
I_{q,C}^{\star,\mu}\approx\frac{1}{2\delta^{2}}\left[1-\frac{B\left(\frac{1}{2};\frac{5}{6}\right)^{3}}{3\pi}2^{-2N_{B}+1}\right].
\label{eq28}
\end{equation}

\subsubsection*{Results for the Gaussian and Cauchy distributions} the validity of the asymptotic approximations will be verified through the evaluation of the FI (\ref{eq4}) for Gaussian and Cauchy distributions with $ \delta=1 $ and for 
\begin{itemize}
\item the optimal set of thresholds for $ N_{B}=\left\lbrace 1,2,3\right\rbrace  $. These optimal thresholds are obtained through exhaustive search. For $ N_{B}=\left\lbrace 4,5,6,7,8\right\rbrace  $ the $ I_{q}^{\star,\mu} $ (\ref{eq26}) and (\ref{eq28}) can be used as approximations.
\item uniform quantization when $ N_{B}=\left\lbrace 1,\cdots,8\right\rbrace  $. In this case, the central threshold is set to $ \mu $ and the optimal quantization interval length $ \Delta^{\star} $ is found by maximizing the FI also using exhaustive search.
\item the practical approximations of the asymptotically optimal thresholds given by (\ref{eq_tau_mu_ggd}) with $ \beta=2 $ and (\ref{eq_tau_mu_cauchy}). $ N_{B}=\left\lbrace 1,\cdots,8\right\rbrace  $ are also considered in this case.
\end{itemize}
The results are given in Tab. \ref{tab1}.

We can verify that in all cases $ I_{q} $ converges fast to $ I_{c} $ when $ N_{B} $ increases. For estimation purposes, 4 quantization bits are enough. Notice that the difference of performance between uniform and non-uniform quantization seems to be slightly higher for the Cauchy distribution. For the Gaussian distribution this difference is negligible, thus in practice, uniform quantization should be used, as it requires lower complexity for implementation. We can also observe that the asymptotic approximation of the FI $ I_{q}^{\star,\mu} $ and its exact value for the practical approximation of the optimal thresholds are close, even when low to medium resolution is used $ N_{B}=4 $ and $ 5 $. At least for $ N_{B}\geq 4 $, this validates the asymptotic approximation of $ I_{q} $ given by (\ref{eq20}) and the approximation of the optimal thresholds (\ref{eq29}) as answers to the questions raised at the beginning of Sec. \ref{sec:p_state}.

\subsection{Scale parameter estimation}
\label{subsec:sca_par_est}

\subsubsection*{Generalized Gaussian distributions} differently from the location problem, the derivative  of the score function exists for all positive $ \beta $. The optimal density is given by
\begin{equation}
\lambda_{GGD}^{\star,\delta}\left(y\right) = \frac{\left\vert \frac{y-\mu}{\delta} \right\vert^{\frac{2\beta-2}{3}}\exp\left(-\frac{1}{3}\left\vert \frac{y-\mu}{\delta} \right\vert^{\beta} \right) }{C_{GGD}^{\delta}}.
\label{eq_lambda_ggd_mu}
\end{equation}
Integrating, we obtain
\begin{equation}
F_{\lambda,GGD}^{\delta}\left(y\right)  = \frac{1}{2}+\frac{\sign\left(y-\mu\right) }{2}\frac{\gamma\left[\frac{1}{3}\left(2+\frac{1}{\beta}\right) ,\frac{1}{3}\left\vert \frac{y-\mu}{\delta} \right\vert^{\beta} \right]}{\Gamma\left[\frac{1}{3}\left(2+\frac{1}{\beta}\right) \right]}.\nonumber
\end{equation}
\noindent Its inverse gives the threshold approximation. For $ i\in\left\lbrace 1,\;\cdots,\;N_{I} \right\rbrace  $
\begin{equation}
\tau_{i,GGD}^{\star,\delta}=\mu+\delta\sign\left(\frac{2i}{N_{I}}-1\right) \alpha_{GGD}^{\delta}\left(\left\vert \frac{2i}{N_{I}}-1 \right\vert\right),
\label{eq_tau_ggd_delta}
\end{equation}
\noindent with
\begin{equation}
\alpha_{GGD}^{\delta}\left(x\right)=\left\lbrace 3\gamma^{-1}\left\lbrace  \frac{1}{3}\left(2+\frac{1}{\beta}\right),x\Gamma\left[\frac{1}{3}\left(2+\frac{1}{\beta}\right) \right]\right\rbrace \right\rbrace^{\frac{1}{\beta}}.
\label{eq_alpha_ggd_delta}
\end{equation}

Observe that all distributions for $ \beta\geq 1 $ have zero density around $ \mu $, which means that $ \mu $ is not the most informative point. This can be understood intuitively by looking for the optimal binary quantizer for symmetric distributions, in this case, the optimal threshold position cannot be $ \mu $, as no information about the density can be obtained from the binary histogram (asymptotically both histogram bins have probability $ \frac{1}{2} $, which are independent of $ \delta $).

The FI for continuous measurements is $ I_{c,GGD}^{\delta}=\frac{\beta}{\delta^{2}} $. After evaluation of the integral in the approximation (\ref{eq22}), we obtain 
\begin{equation}
I_{q,GGD}^{\star,\delta}\approx \frac{\beta}{\delta^{2}}\left[1-2^{-2N_{B}} 3^{1+\frac{1}{\beta}}\beta \frac{\Gamma^{3}\left[\frac{1}{3}\left(2+\frac{1}{\beta} \right)\right]}{\Gamma\left(\frac{1}{\beta}\right) }  \right]  .
\label{eq_iq_ggd_delta}
\end{equation}

In the Gaussian case this FI approximation is 
\begin{equation}
 I_{q,G}^{\star,\delta}\approx \frac{2}{\delta^{2}}\left[1-2^{-2N_{B}+1}\sqrt{\frac{27}{\pi}}\Gamma^{3}\left(\frac{5}{6}\right)  \right].
\label{eq_iq_gaussian_delta}
\end{equation}

\subsubsection*{Student t distributions}
The interval density is the following:
\begin{equation}
\lambda_{STD}^{\delta}\left(y\right) =\frac{1}{C_{STD}^{\delta}}\left\lbrace\frac{\frac{1}{\beta}\left(\frac{y-\mu}{\delta}\right) }{\left[1+\frac{1}{\beta}\left(\frac{y-\mu}{\delta}\right)^{2} \right]^{2}}  \right\rbrace.
\label{eq_lambda_std_mu}
\end{equation}

Now, differently from location parameter estimation, we can evaluate analytically the CDF. Exploiting the symmetry around $ \mu $ and using a change of variables $ \varepsilon=\frac{1}{\left[1+\frac{1}{\beta}\left(\frac{y-\mu}{\delta}\right)^{2} \right]} $, we have
\begin{eqnarray}
\hspace{-90pt}F_{\lambda,STD}^{\delta}\left(y\right)  = \frac{1}{2}+\frac{\sign\left(y-\mu\right) }{2}\times\nonumber\\
\frac{\left[ B\left(\frac{5}{6},\frac{\beta+4}{6} \right)-I_{\frac{1}{1+\frac{1}{\beta}\left(\frac{y-\mu}{\delta}\right)^{2}}}\left(\frac{5}{6},\frac{\beta+4}{6} \right)\right]  }{B\left(\frac{5}{6},\frac{\beta+4}{6} \right)}.\hspace{-80pt}\nonumber
\end{eqnarray}
\noindent For $ i\in\left\lbrace 1,\;\cdots,\;N_{I} \right\rbrace  $, the approximately optimal thresholds are then given by
\begin{equation}
\tau_{i,STD}^{\star,\delta}=\mu+\delta\sign\left(\frac{2i}{N_{I}}-1\right)\alpha_{STD}^{\delta}\left(\left\vert \frac{2i}{N_{I}}-1 \right\vert\right)  .
\label{eq_tau_std_delta}
\end{equation}
\begin{equation}
\alpha_{STD}^{\delta}\left(x\right)=\left\lbrace \beta\left\lbrace I_{B\left(\frac{5}{6},\frac{\beta+4}{6} \right)\left(1-x\right)  }^{-1}\left(\frac{5}{6},\frac{\beta+4}{6} \right)\right\rbrace^{-1}-\beta \right\rbrace^{\frac{1}{2}}.
\label{eq_alpha_std_delta}
\end{equation}

 \begin{table*}[ht]
\centering
\begin{tabular}{ c || c c c || c c c}
 \multicolumn{1}{c}{} & \multicolumn{3}{c}{Gaussian ($ I_{c}=2 $)}&\multicolumn{3}{c}{Cauchy ($ I_{c}=0.5 $)}\\\hline
$ N_{B}  $&Optimal&Uniform&\parbox[c]{1.2cm}{Practical\\approx.}&Optimal&Uniform&\parbox[c]{1.2cm}{\vspace{2pt}Practical\\approx.\vspace{2pt}}\\\hline\hline
$ 1 $&$ 0.60841793^{\dagger}$&$ 0.60841793 $&$ 0 $&$ 0.14332792^{\dagger} $&$ 0.14332792 $&$ 0 $\\\hline
$ 2 $&$ 1.30448971^{\ddag} $&$ 1.30448971 $&$ 1.21760862 $&$ 0.40528473^{\dagger} $&$ 0.40528473 $&$ 0.40528473 $\\\hline
$ 3 $&$ 1.78857963^{\ddag} $&$ 1.77385323 $&$ 1.76989629 $&$ 0.47900864^{\dagger} $&$ 0.47612428 $&$ 0.47893785 $\\\hline
$ 4 $&$ 1.93411857^{\star} $&$ 1.93333156 $&$ 1.93825610 $&$ 0.49533850^{\star} $&$ 0.49213193 $&$ 0.49504170 $\\\hline
$ 5 $&$ 1.98352964^{\star} $&$ 1.98079790$&$ 1.98404286 $&$ 0.49883463^{\star} $&$ 0.49721135 $&$ 0.49879785 $\\\hline
$ 6 $&$ 1.99588241^{\star} $&$ 1.99450778 $&$ 1.99594618 $&$ 0.49970866^{\star} $&$ 0.49898308 $&$ 0.49970408 $\\\hline
$ 7 $&$ 1.99897060^{\star} $&$ 1.99843923 $&$ 1.998978541 $&$ 0.49992716^{\star} $&$ 0.49962443 $&$ 0.49992659 $\\\hline
$ 8 $&$ 1.99974265^{\star} $&$ 1.99956006 $&$ 1.99974364 $&$ 0.49998179^{\star} $&$ 0.49986056 $&$ 0.49998172 $\\\hline
\end{tabular}
\caption{Fisher information (FI) for the estimation of Gaussian and Cauchy scale parameters based on quantized measurements. $ N_{B} $ is the number of quantization bits. In $ \textit{Optimal}^{\dagger} $ the maximum FI obtained by exhaustive search of the optimal threshold is presented, whereas in $ \textit{Optimal}^{\ddag}  $ the optimal quantizer is constrained to be symmetric (with central threshold on $ \mu $). $ \textit{Optimal}^{\star} $ is the theoretical asymptotic approximation of the FI. \textit{Uniform} shows the value of the FI for optimal uniform quantization, the central threshold value is equal to $ \mu $ for $ N_{B}>1 $, for $ N_{B}=1 $ the optimal asymmetric quantizer is considered. \textit{Practical approx.} gives the FI for the practical approximation of the asymptotically optimal thresholds. }
\label{tab2}
\end{table*}

Using the fact that $ I_{c,STD}^{\delta}=\frac{1}{\delta^{2}}\left[3\frac{\beta+1}{\beta+3}-1 \right]  $ and evaluating the integral in (\ref{eq22}), we get
\begin{equation}
I_{q,STD}^{\star,\delta}\approx \frac{1}{\delta^{2}}\left[ 3\frac{\beta+1}{\beta+3}-1-2^{-2N_{B}}\frac{\left(\beta+1\right)^{2}B^{3}\left(\frac{5}{6},\frac{\beta+4}{6} \right) }{3B\left(\frac{1}{2},\frac{\beta}{2} \right)}\right]
\label{eq_iq_std_delta}
\end{equation}
\noindent where in the Cauchy case, we have
\begin{equation}
I_{q,C}^{\star,\delta}\approx \frac{1}{\delta^{2}}\left\lbrace  \frac{1}{2}-2^{-2N_{B}}\frac{\sqrt{\pi} }{3}\left[ \frac{\Gamma\left(\frac{5}{6}\right) }{\Gamma\left(\frac{4}{3}\right)}\right]^{3}\right\rbrace .
\label{eq_iq_cauchy_delta}
\end{equation}

\subsubsection*{Results for the Gaussian and Cauchy distributions}
similarly to the location case we will test the validity of the approximations in the Gaussian and Cauchy cases. The only difference is that we impose the central threshold to be placed at $ \mu $ for $ N_{B}>1 $ when we search for the optimal non-uniform and uniform thresholds. This constraint is used to reduce the search space of thresholds possibilities. Note that this can lead to a possibly suboptimal solution for small numbers of bits ($ N_{B}=2 $) as for $ N_{B}=1 $ the optimal threshold value is not $ \mu $. For $ N_{B}=1 $ we let the optimal quantizer to be asymmetric (with non equiprobable outputs).

The results of the FI evaluations are shown in Tab. \ref{tab2}. 

Similar conclusions to the location estimation case can be drawn: the $ I_{q}^{\star,\delta} $ converges quickly to $ I_{c} $, 4 or 5 bits are enough for estimation of $ \delta $ and for $ N_{B}\geq 4 $, the practical approximation of the optimal thresholds gives a FI close to the asymptotic approximation. The main differences w.r.t. location estimation are obtained for $ N_{B}=1 $, where optimal quantization gives a poor performance. Roughly, we get an asymptotic estimation variance $ 3 $ times higher than in the continuous case. In practice, when $ N_{B}=1 $ we need not only to avoid placing the threshold far from $ \mu $, which is also a regular requirement when we estimate $ \mu $ (see \cite{Papadopoulos2001} and \cite{Ribeiro2006a}), but also avoid placing it near $ \mu $, which is the reason for the zero FI for the practical approximation.

\section{The adaptive approach}
An issue for evaluating the thresholds $ \tau_{i}^{\star} $ (\ref{eq29}) is that they depend on the parameter to be estimated $ x $. To solve this problem, we can initially set $ \tau_{i}^{\star} $ with a guess on $ x $, then estimate $ x $, for example with a MLE, using a set of measurements and update the thresholds with the estimate. We can implement this procedure in a recursive way to obtain a set of thresholds that gets closer and closer to the practical approximation of the optimal set.

\subsection{Adaptive estimation of location}
\label{sec:loc_par_adapt}

In the location estimation case \cite{Papadopoulos2001} and \cite{Fang2008} proposed the use of recursive maximum likelihood estimation for jointly setting the thresholds and estimating the parameter $ \mu $. Note that for a symmetric interval density parametrized with a known $ \delta $ and the unknown $ \mu $, the practical approximation of the thresholds is of the form
\begin{equation}
\tau_{i}^{\star,\mu}=\mu+\delta\sign\left(\frac{2i}{N_{I}}-1\right)\alpha^{\mu}\left(\left\vert \frac{2i}{N_{I}}-1 \right\vert\right)  .
\label{eq_tau_mu}
\end{equation}
Thus the adaptive quantization strategy consists of replacing $ \mu $ with the last MLE $ \hat{\mu}_{k-1}^{MLE} $, which can be easily implemented by subtracting a bias $  \hat{\mu}_{k-1}^{MLE} $ from the input of a quantizer with thresholds
\begin{equation}
{\tau'}_{i}^{\star,\mu}=\delta\sign\left(\frac{2i}{N_{I}}-1\right)\alpha^{\mu}\left(\left\vert \frac{2i}{N_{I}}-1 \right\vert\right).
\label{eq_tau_prime_mu}
\end{equation}
\noindent As $ {\tau'}_{i}^{\star,\mu} $ do not depend on $ \mu $, they represent a static quantizer.

It can be shown that this adaptive estimation/quantization strategy based on the MLE is asymptotically equivalent in terms of performance to an adaptive strategy based on the following low complexity recursive estimator \cite{Farias2013}:
\begin{equation}
\hat{\mu}_{k}=\hat{\mu}_{k-1}+\frac{1}{kI_{q}^{\mu}}\eta_{\mu}\left( i_{k} \right),
\label{eq30}
\end{equation}
\noindent where $ I_{q}^{\mu} $ is the FI for the thresholds $ {\tau'}_{i}^{\star,\mu} $ (which in this case does not depend on $ \mu $). The function $ \eta_{\mu} $ is
\begin{equation}
\eta_{\mu}\left( i \right)=\frac{ f\left({\tau'}_{i-1}^{\star,\mu};\mu=0\right)-f\left({\tau'}_{i}^{\star,\mu};\mu=0\right) }{ F\left({\tau'}_{i}^{\star,\mu};\mu=0\right)-F\left({\tau'}_{i-1}^{\star,\mu};\mu=0\right) }.
\label{eq31}
\end{equation}
\noindent This function is independent of $ \mu $ and if we relate this scheme to standard quantization, the values of $ \eta_{\mu}\left( i \right) $ can be considered as the quantizer output coefficients. The algorithm can be seen either as an asymptotically optimal quantization procedure where the input offset is adjusted as a cumulative function of quantizer output coefficients (cumulative with a decreasing gain to obtain convergence), or as a recursive estimator based on stochastic gradient ascent applied to the log-likelihood ($ \eta_{\mu}\left( i \right) $ is the score function when $ \hat{\mu}_{k}=\mu $).

\begin{figure}[t]
\begin{minipage}[t]{1.0\linewidth}
  \centering
  \centerline{\includegraphics[width=8.5cm]{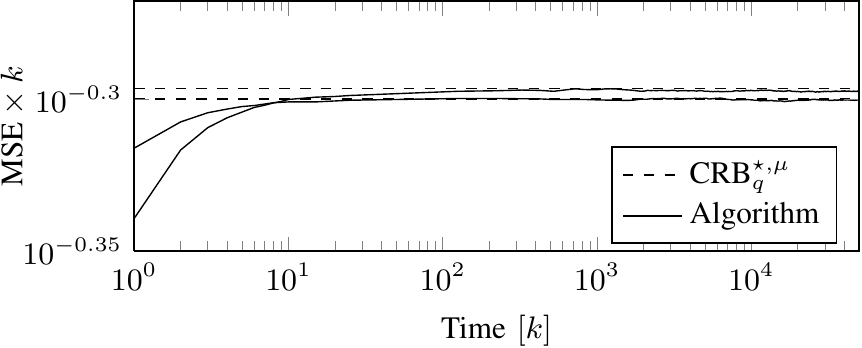}}
  %\vspace{0.3cm}
  \centerline{(a)}
\end{minipage}
\begin{minipage}[t]{1.0\linewidth}
  \centering
  \centerline{\includegraphics[width=8.3cm]{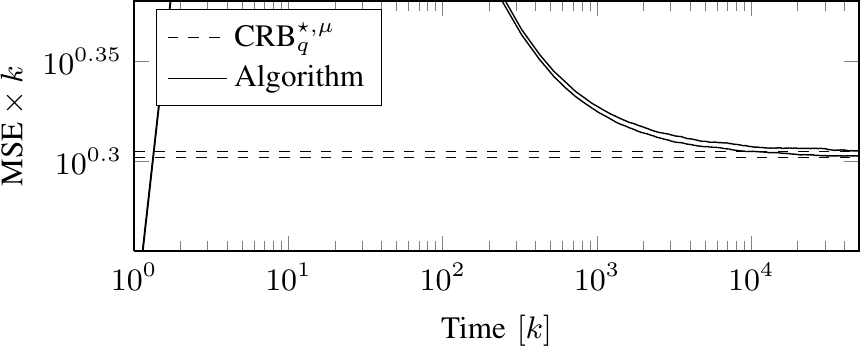}}
%  \vspace{1.5cm}
  \centerline{(b)}
\end{minipage}
\caption{Simulated mean squared error (MSE) for the adaptive estimation of a location parameter (a) Gaussian and (b) Cauchy distributions. The numbers of quantization bits are $ N_{B}=4 $ and $ 5 $. The initial estimation error and $ \delta $  are equal to $ 1 $ in all the cases. The curves that have asymptotically higher values correspond to $ N_{B}=4 $.}
\label{fig1}
\end{figure}

As its performance is equivalent to the MLE performance, if $ N_{B}>4 $, the asymptotic variance of this estimator is close to optimal and it can be approximated with the asymptotic approximation $ I_{q}^{\star,\mu} $
\begin{equation}
\var\left[ \hat{X}_{k}\right] \approx CRB_{q}^{\star,\mu}= \frac{1}{kI_{q}^{\star,\mu}}.
\label{eq32}
\end{equation}
\noindent To validate this approach, we test this algorithm for both Gaussian and Cauchy distributions for $ N_{B}=4 $ and $ 5 $. The estimation mean squared error (MSE) is evaluated through simulation, $ 4\times10^{6} $ realizations of blocks with $ 5\times10^{4} $ samples are simulated. The initial error $ \mu-\hat{\mu}_{0} $ and $ \delta $ are both equal to $ 1 $. The MSE for the algorithm and the approximation given by (\ref{eq32}) are both shown in Fig. \ref{fig1}, where we multiplied them by $ k $ to improve visualization.
We observe that the simulated asymptotic performance is very close to the approximation. For small $ k $ the algorithm is biased and this is a possible reason for the algorithm to perform better than the bound. When comparing this simulation results with uniform quantization simulation results, it was observed that using uniform thresholds leads to faster convergence to the asymptotic performance. This gives some motivation to use an algorithm with changing thresholds. In the initial phase, the convergence phase, a uniform set of thresholds is used, then after a given number of samples, the thresholds change to the practical approximation of the optimal thresholds.

\subsection{Adaptive estimation of scale}
\label{sec:sca_par_adapt}
The practical approximation of the optimal threshold can also be written in a form similar to (\ref{eq_tau_mu}). Thus, after subtracting a fixed known offset $ \mu $ and applying an input gain of the form $ \frac{1}{\hat{\delta}_{k-1}} $, where $ \hat{\delta}_{k} $ is an estimate of $ \delta $, the quantizer can be implemented with thresholds given by
\begin{equation}
{\tau'}_{i}^{\star,\delta}=\sign\left(\frac{2i}{N_{I}}-1\right)\alpha^{\delta}\left(\left\vert \frac{2i}{N_{I}}-1 \right\vert\right).
\label{eq_tau_prime_delta}
\end{equation}
\noindent As $ {\tau'}_{i}^{\star,\delta} $ do not depend on $ \delta $, they also represent a static quantizer.

Similarly to location estimation, we can use a low complexity recursive procedure based on stochastic gradient ascent applied to the log-likelihood to obtain the estimate $ \hat{\delta}_{k} $.
\begin{equation}
\hat{\delta}_{k}=\hat{\delta}_{k-1}+\frac{\hat{\delta}_{k-1}}{kI_{q}^{\delta}}\eta_{\delta}\left( i_{k} \right).
\label{eq30}
\end{equation}
\noindent As the score divided by the FI depends on the unknown $ \delta $ in this case, we replace it by its estimate $ \hat{\delta}_{k-1} $ (the first factor in the second term). $ I_{q}^{\delta} $ is the FI for the thresholds $ {\tau'}_{i}^{\star,\delta} $ when $ \delta=1 $. Now, the output coefficients $ \eta_{\delta} $ are
\begin{equation}
\eta_{\mu}\left( i \right)=\frac{ {\tau'}_{i}^{\star,\delta}f\left({\tau'}_{i}^{\star,\delta};\delta=1\right)-{\tau'}_{i}^{\star,\delta}f\left({\tau'}_{i}^{\star,\delta};\delta=1\right) }{ F\left({\tau'}_{i}^{\star,\delta};\delta=1\right)-F\left({\tau'}_{i-1}^{\star,\delta};\delta=1\right) }.
\label{eq_eta_delta}
\end{equation}
\noindent These coefficients are independent of $ \delta $. In this case, the optimal quantizer is obtained by updating its input gain with a cumulative sum of its output coefficients. The cumulative sum is weighted in this case with a decreasing weight which also depends on the last estimate of the parameter.

\begin{figure}[t]
\begin{minipage}[t]{1.0\linewidth}
  \centering
  \centerline{\includegraphics[width=8.5cm]{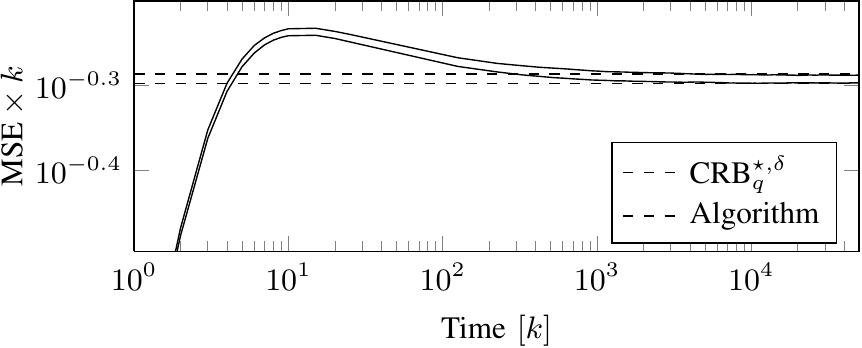}}
  %\vspace{0.3cm}
  \centerline{(a)}
\end{minipage}
\begin{minipage}[t]{1.0\linewidth}
  \centering
  \centerline{\includegraphics[width=8.3cm]{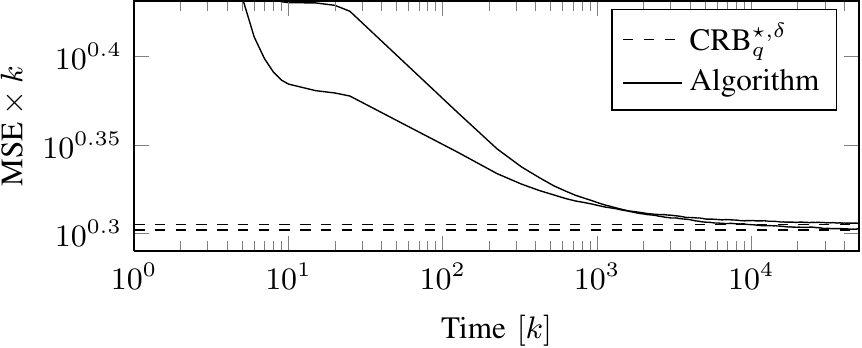}}
%  \vspace{1.5cm}
  \centerline{(b)}
\end{minipage}
\caption{Simulated mean squared error (MSE) for the adaptive estimation of a scale parameter considering (a) Gaussian and (b) Cauchy distributions. The numbers of quantization bits are $ N_{B}=4 $ and $ 5 $. The location parameter is $ \mu=0 $, while the scale parameter value and its initial estimate are $ \delta=1 $ and $ \hat{\delta}_{0}=2 $. The curves that have asymptotically higher values correspond to $ N_{B}=4 $.}
\label{fig2}
\end{figure}

We simulate this algorithm under similar conditions as those used in the location case, the only difference is that $ x=0 $. The initial guess of the parameter is $ \hat{\delta}_{0}=2 $. The results are shown in Fig. \ref{fig2}, where we can see that the simulated performance converges approximately to the theoretical results
\begin{equation}
\var\left[ \hat{X}_{k}\right] \approx CRB_{q}^{\star,\delta}= \frac{1}{kI_{q}^{\star,\delta}}.
\label{eq_var_delta}
\end{equation}

\subsection{Adaptive estimation of location with unknown scale}
In a practical situation we might want to estimate a location parameter when the scale is unknown. In this case, as it is presented in \cite{Farias2013b}, we can blend the two adaptive solutions above, by letting the quantizer to have an adjustable offset given by $ \hat{\mu}_{k-1} $ and an adjustable input gain given by $ \frac{1}{\hat{\delta}_{k-1}} $. Now, the static quantizer has thresholds given by
\begin{equation}
{\tau'}_{i}^{\star,\mu,\delta=1}=\sign\left(\frac{2i}{N_{I}}-1\right)\alpha^{\mu}\left(\left\vert \frac{2i}{N_{I}}-1 \right\vert\right)
\label{eq_tau_prime_mu_delta}
\end{equation}
\noindent and the estimate is given by
 \begin{equation}
\left[\begin{array}{c} \hat{\mu}_{k}\\\hat{\delta}_{k} \end{array} \right]=\left[\begin{array}{c} \hat{\mu}_{k-1}\\\hat{\delta}_{k-1} \end{array} \right]+\frac{\hat{\delta}_{k-1}}{k}\left[ \begin{array}{cc} I_{q}^{\mu,\delta=1 }& 0 \\ 0 & I_{q}^{\delta } \end{array} \right]\left[\begin{array}{c} \eta_{\mu,\delta=1 }\left( i_{k} \right) \\ \eta_{\delta}\left( i_{k} \right) \end{array} \right] 
\label{eq_est_mu_delta}
\end{equation}
\noindent where the superscript $ \delta=1 $ indicates the value of $ \delta $ for which these quantities are evaluated.

If the continuous measurements PDF and the quantizer are symmetric around $ \mu $, which is the case with the practical approximation for the optimal thresholds, then it is possible to show that the asymptotic covariance of estimation of $ \mu $ and $ \delta $ for this algorithm is diagonal with its elements given by the CRB for the separated problems (estimation of $ \mu $ when $ \delta $ is known and of $ \delta $ when $ \mu $ is known) \cite{Farias2013b}. Thus, if $ N_{B}>4 $, we expect that the variance of $ \hat{\mu}_{k} $ is asymptotically given by (\ref{eq32}), even if $ \delta $ is unknown.

\begin{figure}[t]
\begin{minipage}[t]{1.0\linewidth}
  \centering
  \centerline{\includegraphics[width=8.5cm]{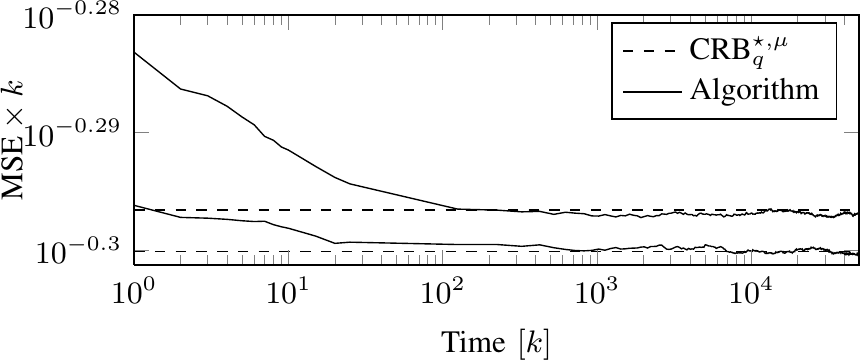}}
  %\vspace{0.3cm}
  \centerline{(a)}
\end{minipage}
\begin{minipage}[t]{1.0\linewidth}
  \centering
  \centerline{\includegraphics[width=8.3cm]{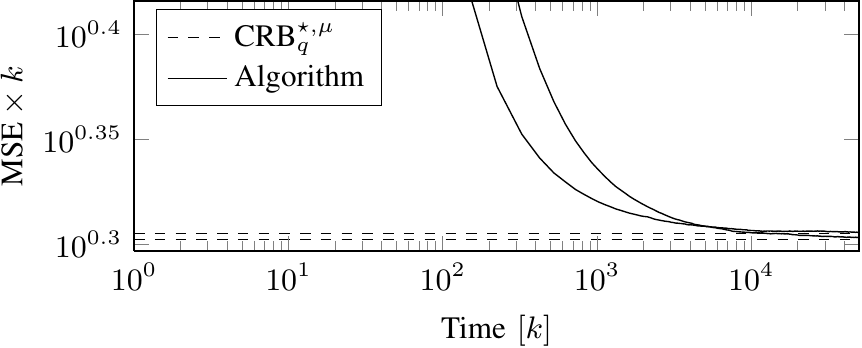}}
%  \vspace{1.5cm}
  \centerline{(b)}
\end{minipage}
\caption{Simulated mean squared error (MSE) for the adaptive estimation of a location parameter of (a) Gaussian and (b) Cauchy distributions when the scale is unknown parameter is unknown. The numbers of quantization bits are $ N_{B}=4 $ and $ 5 $. The value of the parameters are $ \delta=1 $ and $ \mu=0 $ and the initial estimates are $ \hat{\delta}_{0}=1 $ and $ \hat{\mu}_{0}=2 $. The curves that have asymptotically higher values correspond to $ N_{B}=4 $.}
\label{fig3}
\end{figure}

We simulate this algorithm under similar conditions to the simulations performed previously, with $ \delta=1 $, $ \mu=0 $, $ \hat{\mu}_{0}=1 $ and  $ \hat{\delta}_{0}=2 $. The results are shown in Fig. \ref{fig3}. 

The convergence to the asymptotic performance is fast in the Gaussian case, whereas it is slow in the Cauchy case due to the specific form of $ \eta_{\mu,\delta=1 } $ (with a "$ \backsim $" shape) associated with the transient time for the convergence of $ \hat{\delta}_{k} $.

Note that we could also choose to optimally estimate $ \delta $ when $ \mu $ is unknown, this would require to change the thresholds of the static quantizer to
\begin{equation}
{\tau'}_{i}^{\star,\delta=1}=\sign\left(\frac{2i}{N_{I}}-1\right)\alpha^{\delta}\left(\left\vert \frac{2i}{N_{I}}-1 \right\vert\right).
\label{eq_tau_prime_delta_mu}
\end{equation}
\noindent Also, it is possible to maximize a weighted sum of the FI for the two parameters, thus allowing to have a trade off of performance between the parameters. In this case however, the practical approximation of the optimal thresholds is difficult to calculate analytically, even in the Gaussian case.

 \section{Conclusions}
\label{sec:concl}
In this article, we presented an asymptotic approximation of the Fisher information for the estimation of a scalar parameter based on quantized measurements. The approximation is given for a large number of quantization intervals, thus it characterizes estimation performance asymptotically both in terms of number of samples, as the FI is related to estimation performance for large number of samples, and of quantizer resolution. We have shown that the Fisher information loss due to quantization decreases quadratically with the number of quantization intervals, or equivalently, exponentially with the number of quantization bits. This indicates that the best strategy in an estimation context can be based on a low resolution multiple sensor approach. Also, we obtained the optimal quantization interval density and we have shown that it depends not only on $ f^{\frac{1}{3}} $ but also the score function related to the estimation problem.

We applied the results to location and scale parameter estimation for generalized Gaussian distributions and Student t distributions. The application of the results to specific elements of these families of distributions, the Gaussian and Cauchy distributions shows that the asymptotic results are valid for 4 quantization bits or more. As the optimal quantizers can be easily found for 3 or less quantization bits, this result indicates that we can obtain theoretically the optimal thresholds, at least approximately, for all $ N_{B} $.

The comparison with optimal uniform quantization shows that, specially in the Gaussian case, non-uniform quantization is only slightly better. Indicating that if a strong complexity constraint is considered, then, in practice, uniform quantization can be a better solution. 

We show also that the optimal thresholds for estimation depend on the unknown parameter. This motivates the use of an adaptive approach to jointly estimate the parameter and set the quantizer thresholds. We present low complexity adaptive algorithms in the cases of estimation of location and scale parameters, both separately and also jointly. With these algorithms, we show that the asymptotic theoretical results can be approximately achieved in practice.

Multiple extensions of this work can be considered, for example, vector quantization, vector parameters and variable rate quantization (where the output entropy is constrained). Also, we can use the asymptotic approximation of the maximum FI to solve an approximation of the problem of bit allocation in a sensor array that communicates its measurements with a constrained rate.

%\IEEEpubidadjcol

%%%%%%%%%%%%%%%%%%%%%%%%%%%%%%%%%

%\section*{Acknowledgment}
%The authors would like to thank Eric Moisan, Steeve Zozor and Olivier J. J. Michel for their helpful comments and the Erasmus Mundus EBWII program for funding this study.

%%%%%%%%%%%%%%%%%%%%%%%%%%%%%%%%%%%%%%%%%%%%%%%%%%%%%%%%%%%%%%%%%%

\ifCLASSOPTIONcaptionsoff
  \newpage
\fi

\bibliographystyle{IEEEtran}
%\IEEEtriggeratref{4}
\bibliography{biblio}

\vspace{-20pt}
\begin{IEEEbiographynophoto}{Rodrigo Cabral Farias}
was born in Porto Alegre, Brazil, in 1986. He received the B.Sc. degree in electrical engineering from the Federal University of Rio Grande do Sul (UFRGS), Porto Alegre, Brazil, and from the Grenoble Institute of Technology (Grenoble-INP), Grenoble, France, both in 2009. He received the M.Sc and Ph.D. degree in signal processing from Grenoble-INP and Grenoble University in 2009 and 2013 respectively. He is currently working as a post doctoral researcher at the GIPSA-Lab (Grenoble Laboratory of Image, Speech, Signal, and Automation).

His research concerns statistical signal processing, digital communications, sensor networks and tensor methods for signal processing.
\end{IEEEbiographynophoto}
\balance
\begin{IEEEbiographynophoto}{Jean-Marc Brossier}
was born in Thonon, France, in 1965. He received the Ph.D. degree in signal processing in 1992 and the Habilitation a Diriger des Recherches in 2002, both from Grenoble- INP.

He worked as an Assistant Professor for Saint-Etienne University (Universit\'e Jean Monnet) from 1993 to 1995. Since 1995, he has been with Grenoble-INP and GIPSA-Lab. He is now a Professor of electrical engineering and he lectures on signal processing and digital communications. His research interests include statistical signal processing, digital communications, adaptive algorithms and physics.
\end{IEEEbiographynophoto}

\vfill


\end{document}